\documentstyle[osa,manuscript]{revtex}  

\begin{document}                
\title{  Post-Gaussian variational method 
for quantum anharmonic oscillator  }
\author{ Akihiro Ogura }
\address{ Laboratory of Physics \\
College of Science and Technology, Nihon University \\
Funabashi, Chiba 274-8501, Japan  }
\maketitle
\begin{abstract}                
Using the post-Gaussian trial functions, we calculate the variational 
solutions to the quantum-mechanical anharmonic oscillator. 
We evaluate not only the ground state but also some 
excited energies, and compare them with numerical results. 
\end{abstract}

\section{ INTRODUCTION}
\label{INT}
The post-Gaussian trial function provides a robust instrument 
for obtaining the variational solutions~\cite{cooper1,cooper2} .
By using the post-Gaussian trial function, 
one gets an excellent estimate for the ground-state of the system. 
These results indicate the possibility for obtaining better solutions 
for the potential problems, not only the ground state but also 
the excited states. 

In this article, we examine the utility of the post-Gaussian 
trial wave functions.
For this purpose, we employ the anharmonic oscillator 
as an example. 
The anharmonic oscillator has been given the useful laboratory 
to test a nonperturbative approximation.  
We also show explicitly that the post-Gaussian variational method 
gives better solutions to the anharmonic oscillator than 
the ordinary Gaussian one. This may give insight 
into a quantum-field-theoretical problem in the standard 
Gaussian variational approach to 
spontaneous symmetry breakings. 

In the following section, we solve the ground-state energy of 
the anharmonic oscillator using the variational method with 
post-Gaussian trial wave function. In section 3, we calculate 
some excited-state energy levels and compare the results with 
numerical calculations. Section 4 is devoted to discussions. 
Throughout this paper all quantities and variables are assumed to be 
made dimensionless for simplicity. 

\section{  POST-GAUSSIAN VARIATIONAL METHOD AND THE GROUND STATE }
\label{GDS}
The system we consider is the anharmonic oscillator. 
The Hamiltonian is written as follows: 
  \begin{equation}
    H = \frac{p^2}{2} + \frac{x^2}{2}+ g x^{4}  ,
  \label{eq:hamiltonian}
  \end{equation}
where $g$ is a positive constant. 

To obtain the ground-state energy of $H$ using 
the variational method, we employ a post-Gaussian trial function
~\cite{cooper1,cooper2} :
  \begin{equation}
    \psi^{(0)} = {\cal N}^{(0)} \exp \left[ - \frac{\alpha}{2} |x|^{2n} \right] ,
  \label{eq:ground}
  \end{equation} 
where the normalization constant ${\cal N}^{(0)}$ is given by 
  \begin{equation}
    {\cal N}^{(0)} = \sqrt{ \frac{ n \alpha^{ \frac{1}{2n} } }
                                              { \Gamma( \frac{1}{2n} )  } } .
  \end{equation}
$\Gamma(z)$ is Euler's gamma function of argument $z$.
We notice that this trial function in Eq.(\ref{eq:ground}) has two 
variational parameters; 
i.e. $\alpha$ and $n$. 
In the case of $n=1$, this trial function reduces to the familiar 
Gaussian trial function which is often taught 
in an elementary course of quantum mechanics. 
 
Using Eq.(\ref{eq:ground}), we calculate the expectation value 
of the Hamiltonain as follows:
  \begin{eqnarray}
    I^{(0)}(\alpha, n) &\equiv& \langle \psi^{(0)}(n, \alpha)| 
                                                  H | \psi^{(0)}(n, \alpha) \rangle \\
                          &=& \frac{n^{2}}{2} 
                                 \frac{\Gamma( 2-\frac{1}{2n} )}
                                         {\Gamma( \frac{1}{2n} )} 
                                 \alpha^{1/n}
                                 + \frac{1}{2} 
                                    \frac{\Gamma( \frac{3}{2n} )}
                                            {\Gamma( \frac{1}{2n} )} 
                                    \alpha^{-1/n}
                                 + g \frac{\Gamma( \frac{5}{2n} )}
                                               {\Gamma( \frac{1}{2n} )} 
                                      \alpha^{-2/n} . 
  \label{eq:expect}
  \end{eqnarray}
The optimization of Eq.(\ref{eq:expect}) with respect to 
the two parameters $\alpha$ and $n$ 
  \begin{eqnarray}
    \left. \frac{\partial I^{(0)}}
                      {\partial \alpha} \right|_{\alpha_{0}} &=& 0 , \\
    \left. \frac{\partial I^{(0)}}
                     {\partial n} \right|_{n_{0}} &=& 0 ,
  \end{eqnarray} 
can easily 
accomplished numerically. 

In Table 1., we show the numerical result for the ground state. 
The optimal value is found to be at $n=n_{0}=1.13493$ 
that indicates the wave function deviates
from the Gaussian one. 

\section{ THE EXCITED STATES AND THE NUMERICAL RESULTS }
\label{PAA}
The first excited state $\psi^{(1)}$ can be determind 
in such a way 
that the following orthogonality condition is satisfied: 
  \begin{equation}
     < \psi^{(0)}(n_{0}, \alpha_{0}) |  \psi^{(1)}(n_{0}, \beta) > = 0,
  \end{equation}
where $n_{0}$ and $\alpha_{0}$ are fixed values of the variational 
parameters obtained  
in the previous section and $\beta$ is a new variational parameter. 
Then, using this trial function, we optimize the expectation value of 
the Hamiltonian:
  \begin{equation}
    I^{(1)}(\beta, n_{0}) = 
              <\psi^{(1)}(n_{0}, \beta)| H | \psi^{(1)}(n_{0}, \beta)>.
  \label{eq:expect1}
  \end{equation}

To find the second excited state, 
we again optimize the expectation 
value of the Hamiltonian using another trial 
function $\psi^{(2)}$ which is orthogonal to both $\psi^{(0)}$ 
and $\psi^{(1)}$:
   \begin{equation}
     < \psi^{(0)}(n_{0}, \alpha_{0}) |  \psi^{(2)}(n_{0}, \gamma) > = 0,~~
     < \psi^{(1)}(n_{0}, \beta_{0}) |  \psi^{(2)}(n_{0}, \gamma) > = 0 ,
  \end{equation}
where $\beta_{0}$ is the fixed value which make 
Eq.(\ref{eq:expect1}) optimal and $\gamma$ is again 
a new variational parameter.
The same procedure stated above can repeatedly be applied to find 
the higher excited states. 

Now, inspired by the harmonic-oscillator wave functions, 
we propose to take the following 
trial functions for the excited states: 
  \begin{eqnarray}
    \psi^{(1)} &=& {\cal N}^{(1)} x 
                           \exp \left[ - \frac{\beta}{2} |x|^{2n_{0}} \right] ,
                         \hspace{0.5cm}  
                         {\cal N}^{(1)} = 
                            \sqrt{ \frac{ n_{0} \beta^{ \frac{3}{2n_{0}} }  }
                                        { \Gamma( \frac{3}{2n_{0}} ) }  },  
    \label{eq:psi1} \\
    \psi^{(2)} &=& {\cal N}^{(2)} 
                    \left\{ n_{0}(\alpha_{0}+\gamma) |x|^{2n_{0}}-1 \right\}
                    \exp\left[ -\frac{\gamma}{2} |x|^{2n_{0}} \right] , 
                                \nonumber \\ 
       {\cal N}^{(2)} &=& \sqrt{ 
                                 \frac{ 4 n_{0} }
                                      { \Gamma(\frac{1}{2n_{0}}) }
                       \frac{ \gamma^{ 2+\frac{1}{2n_{0}} }  }
                            { (2n_{0}+1) \gamma^{2}
                              +2(2n_{0}-1) \alpha_{0} \gamma 
                              +(2n_{0}+1) \alpha_{0}^{2}      }        }
  \label{eq:psi2}  
  \end{eqnarray}

In Table 1., we compare the results of the post-Gaussian 
trial functions 
with the Gaussian and numerical results~\cite{hioe}. 
As can explicitly be seen, the post-Gaussian trial functions 
always give better 
results than those with the Gaussian trial functions.

\section{ CONCLUSIONS }
We have applied the post-Gaussian trial functions 
to the anharmonic oscillators. 
We have explicitly calculated up to the second excited state and 
compared the results with numerical calculations. 
We have seen how the post-Gaussian trial functions give 
the better results than 
the Gaussian wavefunctions. 

Finally, we note that Cooper et al.~\cite{cooper3,cooper4} have also 
discussed the variational 
energy eigenvalues for the anharmonic oscillator using 
the post-Gaussian trial 
functions. Their algorithm to obtain the excited states is based on 
the idea of SUSY quantum mechanics~\cite{gozzi}, 
whereas the method we 
discussed here is the standard one known in elementary quantum 
mechanics. 

\acknowledgments
This work was begun in collaboration with Dr. S. Abe and I thank him 
for all his contributions.


\newpage

\begin{table}

\caption{ Energy eigenvalues of the anharmonic oscillator with $g=1$ 
in the Gaussian and post-Gaussian variational methods 
in comparison with the numerical results. 
All quantities are dimensionless.  }

\begin{tabular}{|c|c|c|c|}  
                    & Gaussian($n$=1) & post-Gaussian & numerical \\ \hline
  ground        &0.81250&0.80490& 0.80377    \\
                   &($\alpha_{0}$=2.00000)  & $\left(
                                     \begin{array}{@{\,}l}
                                        \alpha_{0}= 1.86647 \\
                                         n_{0}       = 1.13493
                                     \end{array} \right)$ & \\ \hline
  1st excited &  2.75994 & 2.73992          & 2.73789 \\
                    &($\beta_{0}$=2.30891)  
                    & ($\beta_{0}$ = 2.03260) & \\ \hline
  2nd excited  & 5.21980 & 5.20002 & 5.17929 \\
                   &($\gamma_{0}$=2.54205)  & ($\gamma_{0}$= 2.34411 )
                     & \\ 
\end{tabular}
\end{table}

\end{document}